# Quantum Informatics View of Statistical Data Processing

Yu. I. Bogdanov[1][*], N. A. Bogdanova[2]

[1] *Institute of Physics and Technology, Russian Academy of Sciences, Moscow, 117218 Russia*
[2] *National Research University, Moscow Institute of Electronic Technology, Moscow, 124498 Russia*

**Abstract** – Application of root density estimator to problems of statistical data analysis is demonstrated. Four sets of basis functions based on Chebyshev-Hermite, Laguerre, Kravchuk and Charlier polynomials are considered. The sets may be used for numerical analysis in problems of reconstructing statistical distributions by experimental data. Examples of numerical modeling are given.

Quantum informatics is a modern, rapidly evolving field of science and technology, which is based on application of quantum systems for realization of principally new methods of data transmission and computation (quantum communications, quantum cryptography, quantum computers) [1,2].

There are a number of serious obstacles to the construction of efficient systems of quantum information processing. One of them is "fragility" and "intangibility" of a quantum state, which is the main object of quantum physics and quantum informatics. A quantum state is described by a so-called state vector, which is a complex vector in abstract Hilbert space that describes probability amplitudes for observing various basis states. The state vector is a fundamentally different information carrier from its classical analogs. One important distinction of quantum systems from classical ones is the fundamental need for statistical description of their behavior. A measurement upon a quantum object changes its quantum state (wave function reduction). This leads to the necessity for statistical (ensemble) approach: while the act of measurement destroys the quantum state of a single object, there is a whole ensemble of the objects.

A quantum registry, which includes $n$ quantum bits (qubits) is described by a state vector which has $2^n$ complex numbers. Statistically, this implies that control of a quantum system can be described by a multiparametric problem of reconstructing a quantum state of statistical ensemble from measurements upon its representatives.

As a practical matter, the multiparametric problem of reconstruction of quantum states plays an important role in realizing the tasks faced by a developer of quantum informational systems. These interconnected tasks are - generation of quantum systems in given quantum states, their transformation during transmission by quantum channels or during quantum computations and observing (measuring) the output of the system. The ability to reconstruct quantum states provides a basis for solving such problems as fine tuning of quantum informational systems, control of precision and functioning stability, detection of outside intrusions to the system.

---

[*] e-mail: bogdanov@ftian.ru

Multiparametric statistical estimation of quantum states is also clearly interesting in fundamental sense, because it provides a tool for analysis of such basic notions of quantum theory as the statistical nature of its predictions, the superposition principle, Bohr's complementarity principle, etc.

Among all the possible methods of reconstruction of quantum states the most important are those that allow for the precision of estimation close to the theoretically feasible one in problems of high dimension. Construction of such estimators based on traditional methods of mathematical statistics leads to computational difficulties that quickly become insolvable as the dimension of the problem grows. The model which stands out is the so-called root model in which the structure of statistical theory is a-priori consistent with the structure of probabilities in quantum mechanics [3,4].

An application of the root approach to the problems of quantum tomography and quantum cryptography allowed us to experimentally prove in the joint works between Institute of Physics and Technology and Moscow State University the possibility of reconstruction of quantum states with the precision greatly exceeding that in other works [5-9].

The proposed approach is based on a "symbiosis" of mathematical apparatus of quantum mechanics and the Fisher's maximum likelihood principle in order to get multiparameteric asymptotically effective estimation of density as well as quantum states with the most simple and fundamental properties.

The new method is based on representation of probability density as a square of absolute value of some function (called psi-function by analogy with quantum mechanics). The psi-function is presented as a decomposition by an orthonormal set of functions. The coefficients of decomposition are estimated by the maximum likelihood method.

The introduction of a psi-function as a mathematical tool for statistical data analysis implies that instead of density of distribution one considers the square root of it, i.e.:

$$p(x) = |\psi(x)|^2 \qquad (1)$$

It is assumed that the psi-function depends on $s$ unknown parameters $c_0, c_1, \ldots, c_{s-1}$ that are coefficients of decomposition by some set of orthonormal basis functions:

$$\psi(x) = \sum_{i=0}^{s-1} c_i \varphi_i(x) \qquad (2)$$

The maximum likelihood principle implies that the "most likely" estimators of unknown parameters $c_0, c_1, \ldots, c_{s-1}$ are the values that maximize the likelihood function and its logarithm.

$$\ln L = \sum_{k=1}^{n} \ln p(x_k | c) \to \max \qquad (3)$$

Here $x_1, \ldots, x_n$ - is a sample of size $n$.

The condition of extreme value of the logarithmic likelihood function and the normalization condition lead to the following likelihood equation (here it is implied that the basis functions $\varphi_i(x)$ and state vector $c$ are real values):

$$\frac{1}{n}\sum_{k=1}^{n}\left(\frac{\varphi_i(x_k)}{\sum_{j=0}^{s-1}c_j\varphi_j(x_k)}\right) = c_i \qquad i = 0,1,...,s-1 \qquad (4)$$

The likelihood equation has a simple quasi-linear structure and allows for construction of an effective, quickly converging iterative procedure (for instance when the number of estimated parameters lies in tens, hundreds or even thousands) [3,4]. This is how the considered problem differs from other known problems solved by the maximum likelihood method, such as with the growth in the number of estimated parameters the complexity of numerical analysis quickly grows, while the stability of the algorithms steeply falls.

Above we have considered the case of a continuous distribution. The case of a discrete distribution is described similarly.

Let us present some basic sets of basis functions, which are convenient to use for numerical analysis in the problems of reconstruction of statistical distributions by experimental data.

The basis orthonormal set of Chebyshev-Hermite functions has the following form:

$$\varphi_k(x) = \frac{1}{\left(2^k k! \sqrt{\pi}\right)^{1/2}} H_k(x) \exp\left(-\frac{x^2}{2}\right), \qquad k = 0,1,2,... \qquad (5)$$

Here $H_k(x)$ - is the Chebyshev-Hermite polynomial of $k$ - th order.

The basis describes stationary states of a quantum harmonic oscillator.

The Chebyshev-Hermite basis is particularly convenient for the following reason. If in the zero-order approximation the distribution is considered to be Gaussian, then one simply assigns the value of the ground state to the state vector, while deviations from Gaussian distribution are modeled by adding higher order harmonics. Note that the final distribution may greatly differ from the zero-order approximation (it can be asymmetric, multi-modal etc.)

The considered set of basis functions can be effectively and precisely applied to the problems of reconstruction of arbitrary statistical distributions defined on the whole line ($-\infty < x < +\infty$).

For distributions defined on a half-line (say $0 \leq x < +\infty$), it is more convenient to use another basis set of orthonormal functions which is based on Laguerre polynomials. The corresponding basis functions have the following form:

$$\varphi_k(x) = L_k(x)\exp\left(-\frac{x}{2}\right), \qquad k = 0,1,2,... \qquad (6)$$

Here $L_k(x)$ - is Laguerre polynomial of $k$ - th order.

In the considered set of basis functions the ground state corresponds to exponential distribution. Accounting for higher harmonics in the psi-function decomposition describes deviations of the distribution from the exponential one.

The two principal models of discrete distributions of mathematical statistics are binomial and Poisson distributions. In the root approach these distributions act as zero-order approximations for reconstruction of discrete distributions of general form.

An orthonormal set of basis functions based on Kravchuk polynomials allows one to describe binomial type multiparametric distributions

The corresponding discrete distribution is defined at points $x = 0, 1, 2, ..., N$. By analogy to the ordinary binomial distribution, we may call the random number $x$ the number of "successes" in a series of $N$ independent "experiments". At zero-order approximation the distribution under consideration is an ordinary binomial distribution. The considered basis functions are defined as:

$$\varphi_k(x) = \left( \frac{k!(N-k)!}{(pq)^k} \frac{p^x q^{N-x}}{x!(N-x)!} \right)^{1/2} K_k^p(x), \quad k = 0, 1, 2, ..., N \qquad (7)$$

Here $p$ - is the parameter corresponding to the average probability of "success", $q = 1 - p$, $K_k^p(x)$ - Kravchuk polynomial of $k$-th order corresponding to a given $p$.

A basis orthonormal set of functions based on Charlier polynomials allows one to model Poisson-type multiparametric distributions. Corresponding distributions are defined on non-negative whole number points $x = 0, 1, 2, ...$. At zero-order approximation the considered distribution is an ordinary Poisson distribution. Its basis functions have the form:

$$\varphi_k(x) = \left( \frac{\lambda^{k+x} e^{-\lambda}}{k! x!} \right)^{1/2} C_k^\lambda(x), \quad k = 0, 1, 2, ... \qquad (8)$$

Here $\lambda$ - is the parameter corresponding to the average number of "successes" (average value of $x$), $C_k^\lambda(x)$ - is the Charlier polynomial of $k$ - th order corresponding to a given $\lambda$.

To illustrate the method presented above, we present examples of reconstructions of statistical distributions on the following figures. One can note very close correspondence (almost exact) between theoretical distributions and root estimators. For comparison, traditional methods' results are also presented, which clearly lack precision.

The presented research also demonstrates significant superiority of the root approach over kernel estimators of Rosenblatt-Parsen and projection estimators of Chentsov for the problems of reconstruction of statistical distributions.

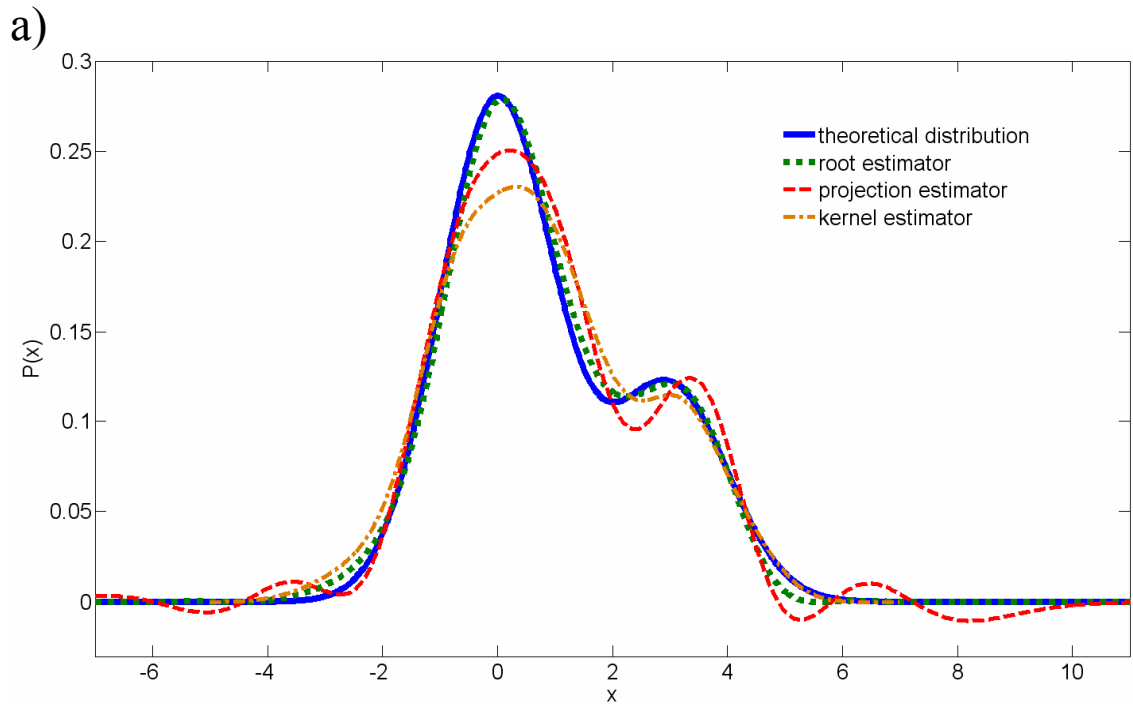

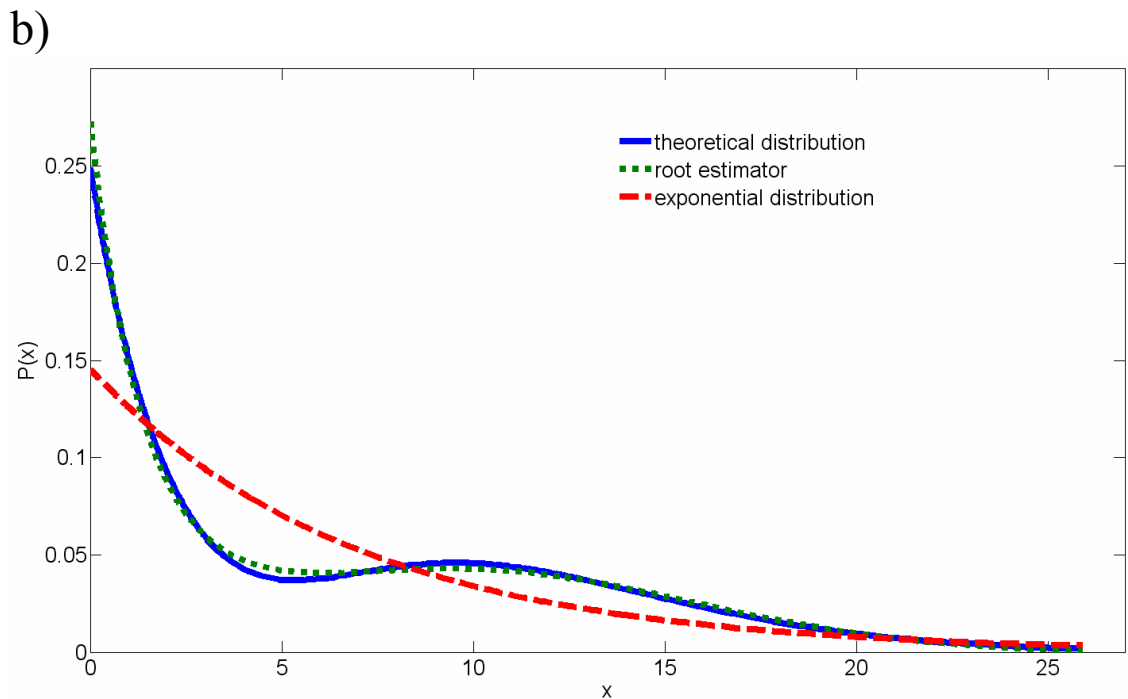

Figure1. Examples of reconstruction of continuous statistical distributions.

a) mixture of two normal distributions with parameters: $\sigma_1 = \sigma_2 = 1$, $\mu_1 = 0$ (70%), $\mu_1 = 3$ (30%). Sample size 200. Comparison of root estimation using Chebyshev-Hermite polynomials basis with kernel and projection estimators.

b) equally- weighted mixture of exponential distribution with mean equal to 2 and chi-squared distribution with 12 degrees of freedom. Sample size 400. Solid line – theoretical density, dot line – root estimation using Laguerre polynomials basis, dashed line – exponential estimation.

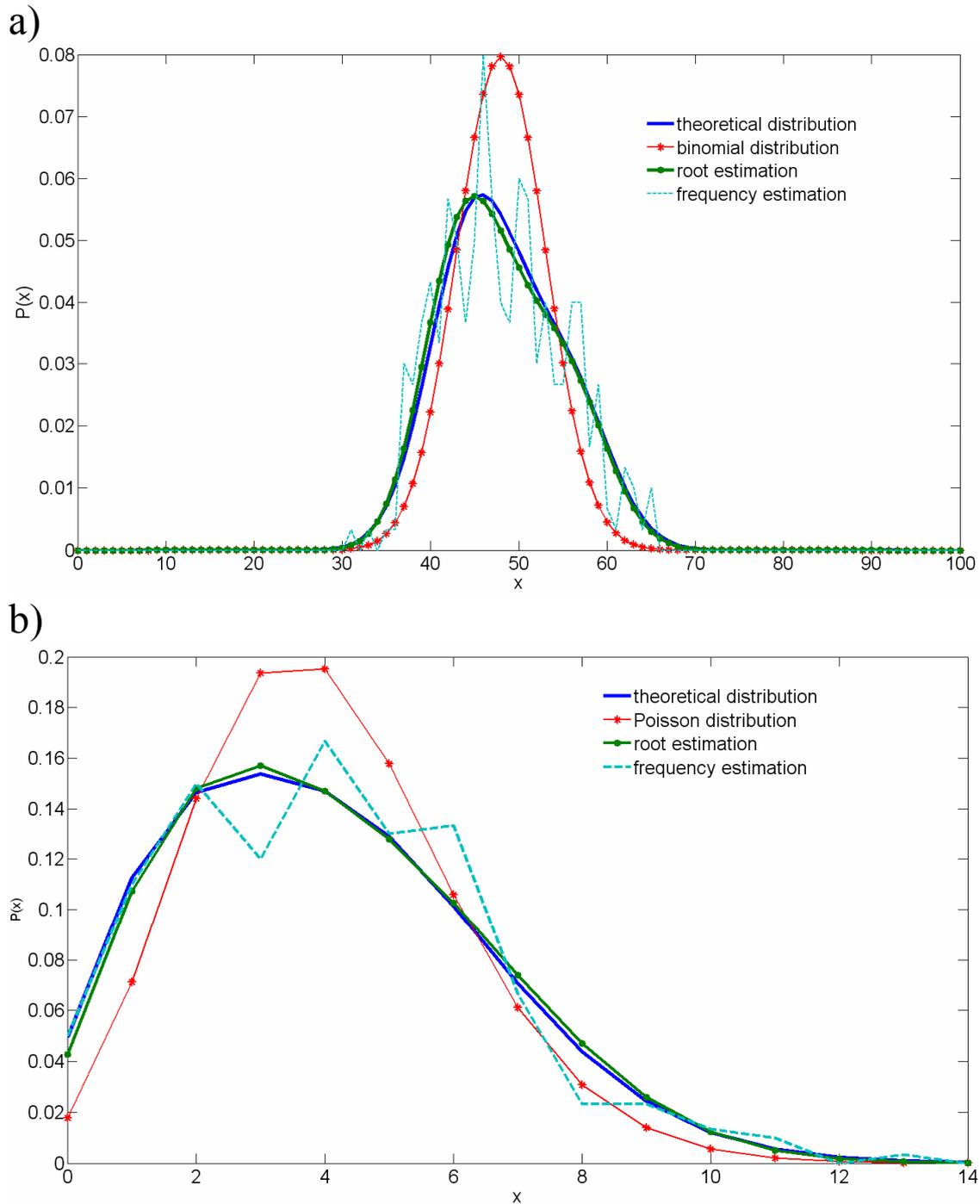

Figure2. Examples of reconstruction of discrete statistical distributions.
a) 2-to-1 weighted mixture of binomial distributions with parameters $N_1 = N_2 = 100$, $p_1 = 0.45$ and $p_2 = 0.55$ correspondingly. Sample size 300. Solid line – theoretical probability distribution, dot line – root estimation with Kravchuk polynomials basis, stars – binomial estimation, dashed line – frequency estimation.
b) 1-to-2 weighted mixture of Poisson distributions with parameters $\lambda_1 = 2$ and $\lambda_2 = 5$ correspondingly. Sample size 300. Solid line – theoretical probability distribution, dot line – root estimation with Charlier polynomials basis, stars – Poisson estimation, dashed line – frequency estimation.